\documentclass[12pt,preprint]{aastex}
\newcommand{\be}{\begin{equation}}
\newcommand{\ee}{\end{equation}}
\newcommand{\bea}{\begin{eqnarray}}
\newcommand{\eea}{\end{eqnarray}}

\begin{document}

\title{Grain Dynamics In Magnetized Interstellar Gas}

\author{A. Lazarian \& Huirong Yan}

\affil{Department of Astronomy, University of Wisconsin, 475 N. Charter
St., Madison, WI53706; lazarian, yan@astro.wisc.edu}

\begin{abstract}
The interstellar medium is turbulent and this induces relative motions
of dust grains. We calculate relative velocities of charged grains
in a partially ionized magnetized gas. We account for anisotropy of
magnetohydrodynamic (MHD) turbulence, grain coupling with magnetic
field, and the turbulence cutoff arising from the ambipolar drag. We obtain
grain velocities for turbulence with parameters consistent with those
in HI and dark clouds. These velocities are smaller than those in
earlier papers, where MHD effects were disregarded. Finally, we consider
grain velocities arising from photoelectric emission, radiation pressure
and H\( _{2} \) thrust. These are still lower than relative velocities induced by turbulence. We conclude that turbulence should prevent these mechanisms from
segregating grains by size. 
\end{abstract}
\section{Introduction}

Dust is an important constituent of the interstellar medium (ISM).
It interferes with observations in the optical range, but provides
an insight to star-formation activity through far-infrared radiation.
It also enables molecular hydrogen formation and traces the magnetic
field via emission and extinction polarization. The basic properties
of dust (optical, alignment etc.) strongly depend on its size distribution.
The latter evolves as the result of grain collisions, whose frequency
and consequences depend on grain relative velocities.

Various processes can affect the velocities of dust grains. Radiation,
ambipolar diffusion, and gravitational sedimentation all can bring
about a dispersion in grain velocities. It was speculated in de Oliveira-Costa
et al. (2000) that starlight radiation can produce the segregation of different
sized grains that is necessary to explain a poor correlation of the
microwave and \( 100\mu m \) signals of the foreground emission (Mukherjee
et al. 2001). If true it has big implications for the CMB foreground
studies. However, the efficiency of this segregation depends on grain
random velocities, which we study in this paper. 

Interstellar gas is turbulent (see Arons \& Max 1975). Turbulence
was invoked by a number of authors (see Kusaka et al. 1970, Volk et
al. 1980, Draine 1985, Ossenkopf 1993, Weidenschilling \& Ruzmaikina
1994) to provide substantial relative motions of dust particles. However, they discussed hydrodynamic turbulence. It is clear
that this picture cannot be applicable to the magnetized
ISM as the magnetic fields substantially affect fluid dynamics.
Moreover dust grains are charged, and their interactions with magnetized
turbulence is very different from the hydrodynamic case. This unsatisfactory
situation motivates us to revisit the problem and calculate the grain
relative motions in magnetized ISM. In what follows, we use the model
of MHD turbulence by Goldreich and Sridhar (1995, henceforth GS95),
which is supported by recent numerical simulations (Cho \& Vishniac
2000, Maron \& Goldreich 2001, Cho, Lazarian \& Vishniac 2002a, henceforth
CLV02). We apply our results to the cold neutral medium (CNM) and a
dark cloud to estimate the efficiency of coagulation, shattering and
segregation of grains.

\section{MHD Turbulence and Grain Motion}

Unlike hydrodynamic turbulence, MHD turbulence is anisotropic, with
eddies elongated along the magnetic field. This happens because it
is easier to mix the magnetic field lines perpendicular to their direction rather than to bend them. The energy of
eddies drops with the decrease of eddy size (e.g. \( v_{l}\sim l^{1/3} \)
for the Kolmogorov turbulence) and it becomes more difficult for smaller
eddies to bend the magnetic field lines. Therefore the eddies get
more and more anisotropic as their sizes decrease. As eddies mix the
magnetic field lines at the rate \( k_{\bot }v_{k} \), where \( k_{\bot } \)is
a wavenumber measured in the direction perpendicular to the local
magnetic field and \( v_{k} \) is the mixing velocity at this scale,
the magnetic perturbations propagate along the magnetic field lines
at the rate \( k_{\parallel }V_{A} \) ,where \( k_{\parallel } \)
is the parallel wavenumber and \( V_{A} \) is the Alfven velocity.
The corner stone of the GS95 model is a critical balance between those
rates, i.e., \( k_{\bot }v_{k} \)\( \sim  \) \( k_{\parallel }V_{A} \),
which may be also viewed as coupling of eddies and wave-like motions.
Mixing motions perpendicular to the magnetic field lines are essentially
hydrodynamic (see CLV02) and therefore it is not surprising that the GS95 predicted
the Kolmogorov one-dimensional energy spectrum in terms of \( k_{\bot } \),
i.e., \( E(k_{\bot })\sim k_{\bot }^{-5/3} \)(see review by Cho, Lazarian \& Yan 2002, henceforth CLY02). 

The GS95 model describes incompressible MHD turbulence. Recent
research suggests that the scaling is approximately true for the dominant Alfvenic modes in a compressible
medium with Mach numbers(\( M\equiv V/C_s \)) of the order of unity (Lithwick \& Goldreich
2001, henceforth LG01, CLY02, Cho \& Lazarian 2002, in preparation), which is also consistent
with the analysis of observational data (Lazarian \& Pogosyan 2000,
Stanimirovic \& Lazarian 2001, CLY02). In what follows we apply the GS95
scaling to handle the problem of grain motions.

Because of turbulence anisotropy, it is convenient to consider separately
grain motions parallel and perpendicular to the magnetic field. The
motions perpendicular to the magnetic field are influenced by
Alfven modes, while those parallel to the magnetic field are subjected
to the magnetosonic modes. The scaling relation for perpendicular motion
is \( v_{k}\propto k_{\perp }^{-1/3} \) (GS95). As the eddy turnover
time is \( \tau _{k}\propto (k_{\perp }v_{k})^{-1} \), the velocity
may be expressed as \( v_{k}\approx v_{max}\left( \tau _{k}/\tau _{max}\right) ^{1/2}, \)
where \( \tau _{max}=l_{max}/v_{max} \) is the time-scale for the
largest eddies, for which we adopt the fiducial values \( l_{max}=10 \)pc,
\( v_{max}=5 \)km/s.

Grains are charged and coupled with the magnetic field. If the Larmor
time \( \tau _{L}=2\pi m_{gr}c/qB \) is shorter than the gas drag
time \( t_{drag} \), grain perpendicular motions are constrained
by the magnetic field. In this case, grains have a velocity dispersion
determined by the turbulence eddy whose turnover period is \( \sim \tau _{L} \), while grains move with the magnetic field on longer time
scales. Since the turbulence velocity grows with the eddy size, the
largest velocity difference occurs on the largest scale where grains
are still decoupled. Thus, following the approach in Draine (1985),
we can estimate the characteristic grain velocity relative to the
fluid as the velocity of the eddy with a turnover time equal to \( \tau _{L} \),
\begin{equation}
\label{vperp}
v_{\perp }(a)={v^{3/2}_{max}\over l^{1/2}_{max}}(\rho _{gr})^{1/2}\left( {8\pi ^{2}c\over 3qB}\right) ^{1/2}a^{3/2},
\end{equation}
 and the relative velocity of grains to each other should be approximately equal to the larger one of the grains' velocities, i.e., the the larger grain's velocity, 
\begin{eqnarray}
\delta v_{\perp }(a_{1},a_{2}) & = & {v^{3/2}_{max}\over l^{1/2}_{max}}(\rho _{gr})^{1/2}\left( {8\pi ^{2}c\over 3qB}\right) ^{1/2}[max(a_{1},a_{2})]^{3/2}\nonumber \\
 & = & 1.4\times 10^{5}cm/s(v_{5}a_{5})^{3/2}/(q_{e}l_{10}B_{\mu })^{1/2},
\end{eqnarray}
 in which \( v_{5}=v_{max}/10^{5} \)cm/s, \( a_{5}=a/10^{-5} \)cm,
\( q_{e}=q/1 \)electron, \( l_{10}=l_{max}/10 \)pc, \( B_{\mu }=B/1\mu  \)G,
and the grain density is assumed to be \( \rho _{gr}=2.6 \)g/cm\( ^{-3} \)
.

Grain motions parallel to the magnetic field are induced by the compressive
component of slow mode with \( v_{\parallel }\propto k_{\parallel }^{-1/2} \)
(CLV02, LG01, CLY02). The eddy turnover time is \( \tau _{k}\propto (v_{\parallel }k_{\parallel })^{-1} \),
so the parallel velocity can be described as \( v_{\parallel }\approx v_{max}\tau _{k}/\tau _{max} \)\footnote{%
We assume that turbulence is driven isotropically at the scale \( l_{max} \).
}. For grain motions parallel to the magnetic field the Larmor precession
is unimportant and the gas-grain coupling takes place on the translational
drag time \( t_{drag} \). The drag time due to collisions with atoms
is essentially the time for collision with the mass of gas equal to
the mass of grain, \( t^{0}_{drag}=(a\rho _{gr}/n)(\pi /8\mu kT)^{1/2}. \), where \( \mu \) is the mass of gas species.
The ion-grain cross-section due to long-range Coulomb force is larger
than the atom-grain cross-section (Draine \& Salpeter 1979). Therefore,
in the presence of collisions with ions, the effective drag time decreases,
\( t_{drag}=\alpha t^{0}_{drag} \), where \( \alpha <1 \) is the
function of a particular ISM phase. The characteristic velocity of
grain motions along the magnetic field is approximately equal to the
parallel turbulent velocity of eddies with turnover time equal to
\( t_{drag} \)

\begin{equation}
\label{vpara}
v_{\parallel }(a)=\alpha {v^{2}_{max}\over l_{max}}\left( \frac{\rho _{gr}}{4n}\right) \left( {2\pi \over \mu kT}\right) ^{1/2}a,
\end{equation}
 and the relative velocity of grains for \( T_{100}=T/100 \)K is

\begin{eqnarray}
\delta v_{\parallel }(a_{1},a_{2}) & = & \alpha {v^{2}_{max}\over l_{max}}\left(\frac{\rho _{gr}}{4n}\right)({2\pi \over \mu kT})^{1/2}[max(a_{1},a_{2})]\nonumber \\
 & = & (1.0\times 10^{6}cm/s)\alpha v_{5}^{2}a_{5}/(nl_{10}T_{100}^{1/2}),
\end{eqnarray}

When \( \tau _{L}>t_{drag} \), grains are no longer tied to the magnetic
field. Since at a given scale, the largest velocity dispersion is
perpendicular to the magnetic field direction, the velocity gradient
over the grain mean free path is maximal in the direction perpendicular
to the magnetic field direction. The corresponding scaling is analogous
to the hydrodynamic case, which was discussed in Draine (1985): \( \delta v(a_{1},a_{2})=v^{3/2}_{max}/ l^{1/2}_{max} t_{drag}^{1/2} \), i.e.,
\begin{equation}
\delta v(a_{1},a_{2})=  \alpha ^{\frac{1}{2}}{v^{3/2}_{max}\over l^{1/2}_{max}}\left( \frac{\rho _{gr}}{4n}\right) ^{\frac{1}{2}}\left( {2\pi \over \mu kT}\right) ^{\frac{1}{4}}[max(a_{1},a_{2})]^{\frac{1}{2}}.\label{HD} 
\end{equation}

Turbulence is damped due to the viscousity when the cascading rate \( v_\perp k_\perp \) equals the damping time \( t_{damp} \)
(see Cho, Lazarian \& Vishniac 2002b). If
the mean free path for a neutral particle \( l_{n} \), in a partially
ionized gas with density \( n_{tot}=n_{n}+n_{i} \), is much less
than the size of the eddy in consideration, i.e., \( l_{n}k_{\bot }\ll 1 \),
the damping time is \( t_{damp}\sim \nu _{n}^{-1}k_{\perp }^{-2}\sim \left( n_{tot}/n_{n}\right) \left( l_{n}v_{n}\right) ^{-1}k_{\perp }^{-2}, \)
where \( \nu _{n} \) is effective viscosity produced by neutrals.
In the present paper we consider cold gas with low ionization, therefore
the influence of ions on \( l_{n} \) is disregarded. Thus the turbulence
cutoff time in neutral medium is
\begin{equation}
\label{cutoff}
\tau _{c}\simeq \left( \frac{l_{n}}{v_{n}}\right) \left( \frac{v_{n}}{v_{max}}\right) ^{\frac{3}{2}}\left( \frac{l_{max}}{l_{n}}\right) ^{\frac{1}{2}}\left( \frac{V_{A}}{v_{max}}\right) ^{\frac{1}{2}}\left( \frac{n_{n}}{n_{tot}}\right) ,
\end{equation}
where \( v_{n} \) and \( V_{A} \) are, respectively, the velocity
of a neutral and Alfven velocity. It is easy to see that for \( \tau _{c} \)
longer than either \( t_{drag} \) or \( \tau _{L} \) the grain motions
get modified. A grain samples only a part of the eddy before gaining
the velocity of the ambient gas. In GS95 picture, the shear rate \( dv/dl \)
increases with the decrease of eddy size. Thus for \( \tau _{c}>max\{t_{drag},\tau _{L}\} \),
these smallest available eddies are the most important for grain acceleration.
Consider first the perpendicular motions. If \( v_{c} \) is the velocity
of the critically damped eddy, the distance traveled by the grain
is \( \bigtriangleup l\sim v_{c}\times min\lbrace t_{drag},\tau _{L}\rbrace  \).
Thus the grain experiences the velocity difference \( \bigtriangleup l\times dv/dl\sim v_{c}\times min\lbrace t_{drag},\tau _{L}\rbrace /\tau _{c} \). Due to the critical balance in GS95 model, the shear rate along the magnetic field is \( dv/dl=v_ck_\parallel\sim v_c/(V_A\tau_c) \). Therefore, grain experiences a velocity difference
\( V_{A}/v_{c} \) times smaller, i.e., \( \sim v_c^2\times t_{drag}/(V_A\tau _{c}) \).

\section{Discussion }

\subsection{Shattering and Coagulation }

Consider the cold neutral medium (CNM) with temperature \( T=100 \)K, density \( n_{\rm H}=30 \)cm\( ^{-3} \),
electron density \( n_{e}=0.045 \)cm\( ^{-3} \), magnetic field
\( B\sim 1.3\times 10^{-5} \)G (Weingartner \& Draine 2001a, hereafter WD01a). To account for the Coulomb drag,
we use the results by WD01a and get the modified drag time \( t_{drag}=\alpha t^{0}_{drag} \).
Using the electric potentials in Weingartner \& Draine (2001b), we
get grain charge and \( \tau _{L} \).

For the parameters given above, we find that \( t_{drag} \) is larger
than \( \tau _{c} \) for grains larger than \( 10^{-6} \)cm, \( \tau _{L} \)
is smaller than \( \tau _{c} \) even for grains as large as \( 10^{-5} \)cm.
Here, we only consider grains larger than \( 10^{-6} \)cm, which
carry most grain mass (\( \sim 80\% \)) in ISM, so we can still use
Eq.(\ref{vpara}) to calculate grain parallel velocities and Eq.(\ref{vperp}) to get the perpendicular velocity for grain larger than \( 10^{-5} \)cm. Nevertheless,
the perpendicular velocities of grains smaller than \( 10^{-5} \)cm should be estimated as \( v'_{\perp }(a)=v_{c}\times (\tau _{L}/\tau _{c})=v_{max}(\tau _{c}/\tau _{max})^{1/2}(\tau _{L}/\tau _{c})=v_{\perp }(a)(\tau _{L}/\tau _{c})^{1/2}, \)
where \( v_{\perp }(a) \) is given by Eq.(\ref{vperp}). The results
are shown in Fig.1.

The critical sticking velocity were calculated in Chokshi et al. (1993)(see also Dominik \& Tielens 1997).\footnote{There are obvious misprints in the numerical coefficient of Eq.(7) in Chokshi et al.(1993) and the power index of Young's modulus in Eq.(28) of Dominik \& Tielens (1997).} However, experimental work by Blum (2000) shows that the critical velocity  is an order of magnitude larger than the theoretical calculation. Thus the collisions can result in coagulation for
small silicate grains (\(\leq 3\times 10^{-6} \)cm). 

With our input parameters, grains do not shatter if the shattering
thresholds for silicate is \( 2.7 \)km/s as in Jones et al. (1996).
Nevertheless, the grain velocities strongly depend on \( v_{max} \)
at the injection scale. For instance, we will get a cutoff \( 6\times 10^{-5} \)cm
due to shattering if \( v_{max}=10 \)km/s.

For a dark cloud, the situation is different. As the density increases,
the drag by gas becomes stronger. Consider a typical dark cloud with
temperature \( T=20 \)K, density \( n_{\rm H}=10^{4} \)cm\( ^{-3} \)
(Chokshi et al. 1993) and magnetic field \( B\sim 2.3\times 10^{-4} \)G.
Assuming that dark clouds are shielded from radiation, grains get charged
by collisions with electrons: \( <q>=0.3(r/10^{-5} \)cm) electrons.
The ionization in the cloud is \( \chi =n_{e}/n_{tot}\sim 10^{-6} \)
and the drag by neutral atoms is dominant. From Eq.(\ref{cutoff}) and
the expression for the drag time and the Larmor time, we find \( \tau _{L}<t_{drag} \)
for grains of sizes between \( 10^{-6} \)cm and \( 4\times 10^{-6} \)cm,
and \( t_{drag}<\tau _{L} \) for grains larger than \( 4\times 10^{-6} \)cm.
In both cases, turbulence cutoff \( \tau _{c} \) is smaller than
\( t_{drag} \) and \( \tau _{L} \). Thus for the smaller grains,
we use Eq.(\ref{vperp}),(\ref{vpara}) to estimate grain velocities.
For larger grains, grain velocities
are given by Eq.(\ref{HD}).

Our results for dark clouds show only a slight difference from the
earlier hydrodynamic estimates. Since the drag time \( t_{drag}\propto n^{-1} \),
Larmor time \( \tau _{L}\propto B^{-1}\propto n^{-1/2} \), the grain
motions get less affected by the magnetic field as the cloud becomes
denser. Thus we agree with Chokshi's et al. (1993) conclusion that
densities well in excess of \( 10^{4} \)cm\( ^{-3} \) are required
for coagulation to occur. Shattering will not happen because the velocities
are small, so there are more large grains in dark clouds. This agrees
with observations (see Mathis 1990).

In the treatment above we disregarded the possibility of direct acceleration
of charged grains through their interactions with fluctuating magnetic
field. In our next paper we will show that this resonant process is important for a highly ionized
medium.

\subsection{Grain Segregation and Turbulent Mixing}

Our results are also relevant to grain segregation. Grains are the
major carrier of heavy elements in the ISM. The issue of grain segregation
may have significant influence on the ISM metallicity. Subjected to
external forcing, e.g., due to radiation pressure, grains gain size-dependent
velocities with respect to gas. WD01a have considered the forces on
dust grains exposed to anisotropic interstellar radiation fields.
They included photoelectric emission, photodesorption as well as radiation
pressure, and calculated the drift velocity for grains of different
sizes. The velocities they got for silicate grains in the CNM range
from \( 0.1 \)cm/s to \( 10^{3} \)cm/s. Fig.1 shows that the turbulence
produces larger velocity dispersions.\footnote{%
If reconnection is fast (see Lazarian \& Vishniac 1999), the mixing
of grains over large scales is provided by turbulent diffusivity\( \sim v_{max}l_{max} \).
On small scales the grain decoupled motions are important.
} Thus the grain segregation of very small and large grains speculated
in de Oliveira-Costa et al. (2000) is unlikely to happen for typical
CNM conditions.

A different mechanism of driving grain motions is a residual imbalance
in {}``rocket thrust{}'' between the opposite surfaces of a rotating
grain (Purcell 1979). This mechanism can provide grain relative motions and preferentially move grains into molecular
clouds. It is easy to see that due to averaging caused by grain rotation,
the rocket thrust is parallel to the rotation
axis. Three causes for the thrust were suggested by Purcell (1979):
spatial variation of the accommodation coefficient for impinging atoms,
photoelectric emission, and H\( _{2} \) formation. The latter was
shown to be the strongest among the three. The uncompensated force
in this case arises from the difference of the number of catalytic
active sites for H\( _{2} \) formation on the opposite grain surfaces.
The nascent H\( _{2} \) molecules leave the active sites with kinetic
energy \( E \) and the grain experiences a push in the opposite
directions. The number of active sites varies from one grain to another,
and we should deal with the expectation value of the force for a given
distribution of active sites.

 Due to internal relaxation of energy (see Lazarian \& Draine
1999a,b, and review by Lazarian 2000) the grain rotational axis tends to be
perpendicular to the largest \( b-b \) surface. Adopting the approach in Lazarian \& Draine (1997), we get the mean
square root force of H\( _{2} \) thrust on a grain in the shape of
a square prism with dimensions \( b\times b\times a \) (\( b>a \))

\begin{equation}
\label{H}
\langle F_{z{\rm H}}\rangle =r^{3/2}(r+1)^{1/2}\gamma (1-y)n_{\rm H}v_{\rm H}a^{2}\left( \frac{2m_{\rm H}E}{\nu }\right) ^{1/2},
\end{equation}
where \( r=b/2a, \), \( n_{\rm H}\equiv n({\rm H})+2n({\rm H}_2) \), \( y=2n({\rm H}_2)/n_{\rm H} \) is the \( {\rm H_2} \) fraction, \( \gamma  \) is the fraction of impinging {\rm H}
atoms and \( \nu  \) is the number of active sites over the grain
surface. The expected grain
velocity is \( v=\langle F_{z{\rm H}}\rangle t_{drag}/m \). In the CNM we consider, \( y=0 \), adopting the characteristic values in Lazarian \& Draine (1997), \( r=1, \)
\( \gamma =0.2 \), \( E=0.2 \)eV, and the density of active sites \( 10^{11} \)cm\(^{-2}\) so that \( \nu =80(a/10^{-5} \)cm\( )^{2}r(r+1) \),
we get the {}``optimistic{}'' velocity shown in Fig 1. For maximal
active site density \( 10^{15} \)cm\( ^{-2} \), we get the lower
boundary of grain velocity \( v\simeq 3.3(10^{-5}{\rm cm} /a)^{1/2} \)cm/s. The scaling is approximate due to the complexity of coefficient \( \alpha \)(see WD01a Fig.16).

Lazarian \& Draine (1999a,b) have shown that subjected to {\rm H}\( _{2} \) torques alone, grains
\( \leq 10^{-4} \)cm should experience frequent thermal flipping,
which means that the \( F_{z{\rm H}} \) fluctuates. This flipping
results from coupling of grain rotational and vibrational degrees
of freedom through internal relaxation and would average out \( \langle F_{z{\rm H}}\rangle  \).
However, the flipping rate depends on the value of the grain angular
momentum (Lazarian \& Draine 1999a). If a grain is already spun up to a sufficient velocity,
it gets immune to thermal flipping. Radiative torques (Draine \& Weingartner
1996) can provide efficient spin if the grain size is comparable
to the wavelength. For a typical interstellar diffuse radiation field,
the radiative torques are expected to spin up grains with sizes larger
than \( \sim 4\times 10^{-6} \)cm. They will also align grains with
rotational axes parallel to the magnetic field. Thus grains should acquire
velocities along the magnetic field lines and the corresponding velocities
should be compared with those arising from turbulent motions parallel to the magnetic
field. It is clear from Fig.1 that for the chosen set of parameters
the effect of {\rm H}\( _{2} \) thrust is limited. All in all, we conclude that the radiation effects and H\(_2\) thrust are not efficient for segregating grains in typical ISM conditions.
\begin{figure}
\centering \leavevmode
\resizebox*{0.33\textwidth}{0.25\textheight}{\includegraphics{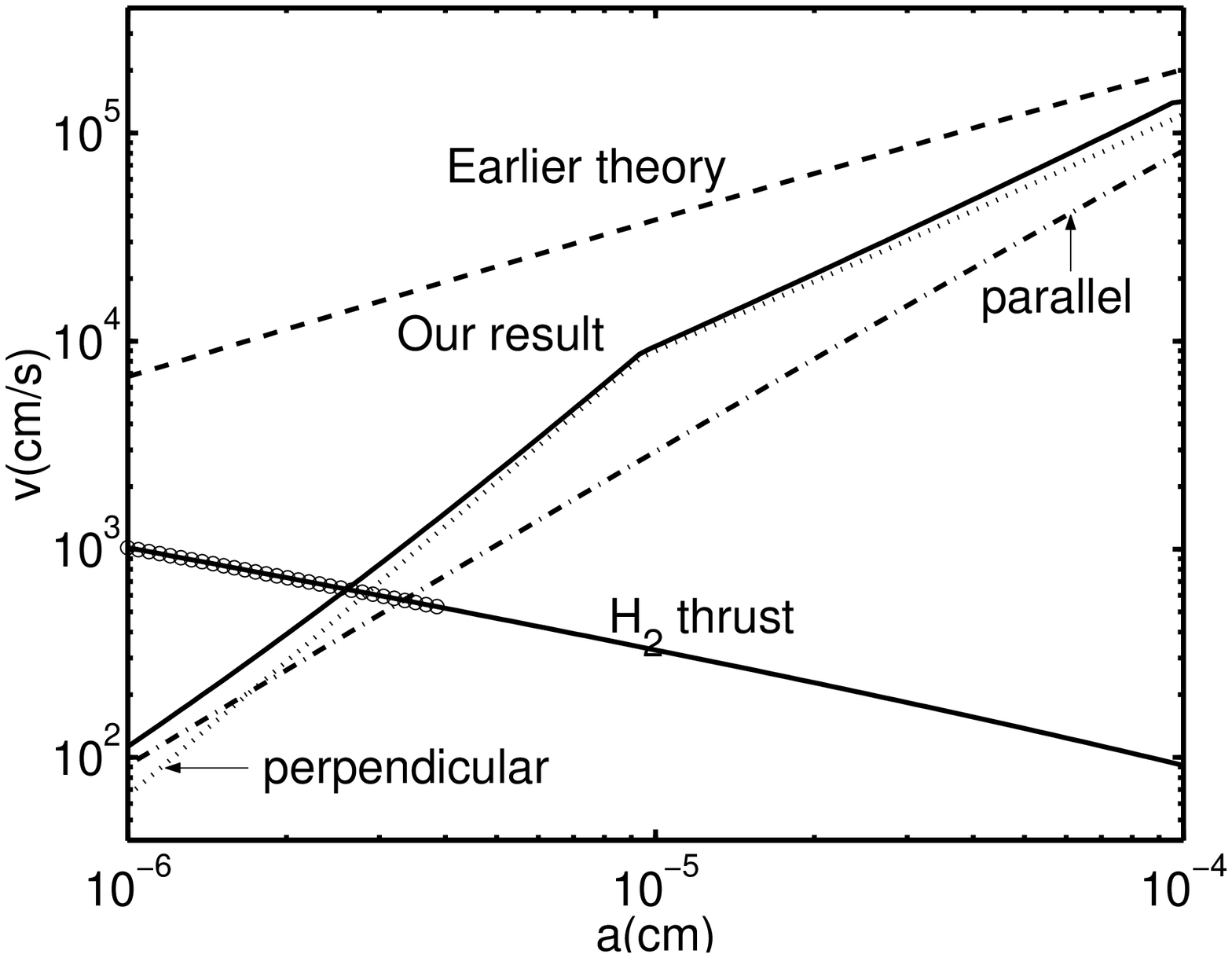}}

\caption{Grain velocities as a function of radii (solid line) in
the CNM. Dashdot line represents parallel velocity
due to the drag by compressible modes, dotted line refers to perpendicular
velocity from the contribution of the drag by Alfven mode, also plotted
is the earlier hydrodynamic result (dashed line). The change of the slope
is due to the cutoff of turbulence by ambipolar diffusion. The grain
velocity driven by {\rm H}\protect\( _{2}\protect \) thrust is plotted
to illustrate the issue of grain segregation in the CNM
(see text), the part marked by 'o' is nonphysical because thermal
flipping is not taken into account. }
\end{figure}

\section{Summary}

We have calculated relative motions of dust grains in a magnetized turbulent
fluid taking into account turbulence anisotropy, turbulence damping
and grain coupling with the magnetic field. We find that these effects
decrease the relative velocities of dust grains compared to the earlier
hydrodynamic-based calculations. The difference is substantial in
CNM, but less important for dark clouds. For CNM we find that coagulations
of silicate grains happen for sizes \(\leq 3\times 10^{-6} \)cm. The
force due to {\rm H}\( _{2} \) formation on grain surface might drive small
grains (\( <3\times 10^{-6} \)cm) to larger velocities but thermal
flipping of grains suppresses the forces for grains less than \( 4\times 10^{-6} \)cm.
We conclude that radiation and H\( _2 \) thrust are not capable of segregating grains. 

We are grateful to John Mathis for reading the manuscript and many
important comments. We thank our referee Dr. Stuart Weidenschilling for helpful comments. The research is supported by the NSF grant AST0125544.

\end{document}